**Evaluating WAIC and PSIS-LOO for Bayesian Diagnostic Classification Model Selection**

Ae Kyong Jung and Jonathan Templin

University of Iowa

**Author Note**

Ae Kyong Jung https://orcid.org/0000-0003-0665-4876

Jonathan Templin https://orcid.org/0000-0001-7616-0973

Email: aekyong-jung@uiowa.edu, jonathan-templin@uiowa.edu

**Abstract**

Bayesian diagnostic classification models (Bayesian DCMs) are effective for diagnosing students' skills. Research on the evaluation of relative model fit indices for DCMs using Bayesian estimation, however, is deficient. This study introduces the performance of Bayesian relative model fit indices, the widely applicable information criterion (WAIC) and leave-one-out cross-validation using Pareto-smoothed importance sampling (PSIS-LOO), in comparison to simpler and more widely used deviance information criterion (DIC). The simulation study evaluates the performance of WAIC and PSIS-LOO by detecting the true model with varying sample sizes, item qualities, and prior information levels. The results of the study indicate that WAIC and PSIS-LOO primarily favored the generating model; however, occasional inconsistencies were observed. This study recommends using WAIC and PSIS-LOO when the data is assumed to follow a simpler model and the models are estimated under uninformative priors, and DIC when the data is assumed to follow a more complex model.

**Introduction**

Diagnostic Classification Models (DCMs) are useful methods for diagnosing student's fine-grained skills or attributes (Rupp et al., 2010). With many diagnostic classification models in use, relative model fit methods play a significant role in determining which model provides the best explanation for the data. Selecting the most appropriate model that fits the data enables a more accurate estimation of respondents' attribute statuses. Conversely, if an ill-fitting model was selected for the data, the estimates of the model cannot be meaningfully interpreted. Despite the extensive use of Bayesian DCMs, research on evaluating the performance of relative model fit indices for these models remains limited. To address the issue, this paper evaluates the performance of relative model fit indices, Widely Applicable Information Criterion or Watanabe-Akaike Information Criterion (WAIC; Watanabe, 2010), and Leave-One-Out (LOO; Vehtari et al., 2017), to determine the best-fitting Bayesian Diagnostic Classification Models for the data.

DCM research has been conducted to assess the performance of model fit indices that can indicate the most appropriate model for the given data. When the model parameters are estimated using Maximum likelihood (ML), the performance of various indices, such as the likelihood ratio test (LRT), the Akaike's Information Criterion (AIC; Akaike, 1974), the Bayesian Information Criterion (BIC; Schwarz, 1978), and the consistent AIC (Bozdogan, 1987) have been investigated (e.g., Hu et al., 2016; Jiao, 2009; Lei & Li, 2016; Ravand & Robitzsch, 2018; Sen & Bradshaw, 2017; Yamaguchi & Okada, 2018).

Research evaluating the performance of relative model fit indices for DCMs using Bayesian inference, however, has not been conducted extensively, even if several indices have been utilized in studies (e.g., Chen et al., 2018; Culpepper & Hudson, 2018; De La Torre & Douglas, 2008; Huang & Wang, 2014; Sinharay & Almond, 2007; Zhan et al., 2019). Sinharay

& Almond (2007) compared the predictive accuracy between a three-class latent class model and a two-class latent class model using the deviance information criterion, (DIC; Spiegelhalter et al., 2002) which is a partially Bayesian index, and the DIC successfully indicated an improvement in model fit when transitioning from the two-class latent class model to the three-class latent class model. De la Torre & Douglas (2008) employed the log marginal likelihood, AIC, BIC and DIC as model fit indices, and they found that all indices selected the same DCM as the best model for real data. However, they argued that marginal likelihood used in these indices to represent the model fit had limitations when applied to Bayesian DCMs, as the marginal likelihood treats the structural parameters, which are the focus of DCMs, as missing data. Also, since AIC, BIC and DIC utilize point estimates instead of the posterior distribution, these indices restrict the utilization of the complete information from the posterior distribution (Luo & Al-Harbi, 2017).

  New methods for selecting Bayesian model fit have been proposed to address the limitations of existing approaches (e.g., Vehtari et al., 2017). The WAIC and LOO make use of the entire posterior distribution. WAIC is considered an improvement over DIC as it computes the penalty term as the variance across the parameters. In other words, the key difference lies in the fact that the penalty term of the WAIC is computed over all individual data points (Vehtari et al., 2017). LOO is a cross-validation technique that involves fitting the model after excluding one observation at a time. As LOO can be computationally intensive, importance sampling (e.g., Gelfand et al., 1992) is utilized to approximate refitting of the model without repetitive computations. Furthermore, Pareto-smoothed importance sampling (PSIS-LOO; Vehtari et al., 2015) has been proposed as a method to stabilize the importance sampling and enhance the accuracy of LOO estimate.

WAIC and LOO have been evaluated in multidimensional IRT and latent class model applications (AlHakmani & Sheng, 2023; Fujimoto & Falk, 2022; Li et al., 2016; Luo & Al-Harbi, 2017; Revuelta & Ximénez, 2017; Tong et al., 2022). AlHakmani & Sheng (2023) conducted an evaluation of WAIC and PSIS-LOO for a mixture Item Response Theory (IRT) model, successfully identifying the correct number of latent classes using the effective number of parameters. Fujimoto & Falk (2022) compared non-nested-dimensionality IRT and nested-dimensionality IRT (such as the bifactor model and two-tier model) using DIC, WAIC, PSIS-LOO and log-predicted marginal likelihood. DIC strongly favored nested-dimensionality IRT over non-nested-dimensionality IRT which was the correct model, and it also tended to favor more complex models within the non-nested-dimensionality IRT framework. In contrast, PSIS-LOO exhibited the least bias, followed by WAIC and log-predicted marginal likelihood. Luo & Al-Harbi (2017) approximated the performance of WAIC, PSIS-LOO, LRT, DIC, AIC, and BIC for 1PL, 2PL and 3PL IRT models. WAIC and PSIS-LOO outperformed the other indices for 3PL model, performed equally well with the others for the 2PL models, and performed worse than the others for the 1PL model.

The objective of this study is to assess the effectiveness of fully Bayesian information criteria, specifically WAIC and PSIS-LOO, in determining the appropriate DCMs estimated using Markov Chain Monte Carlo (MCMC) algorithms. The findings of this research will provide valuable insights for researchers and practitioners utilizing Bayesian DCMs. To facilitate a comparison with partially Bayesian information criteria, DIC was included in our study. For generating and estimating models, we employed the log-linear cognitive diagnostic model (LCDM; Henson et al., 2009), along with its nested model variants, the deterministic inputs, noisy-and-gate (DINA; Haertel, 1989; Junker & Sijtsma, 2001) model and the compensatory reparameterized unified model (CRUM; Hartz, 2002). The inclusion of two variants allows us to compare which of the three model fit indices perform better in which model. All three models are equivalent when items measure only one attribute. We first investigated the parameter recovery in DCMs using MCMC to ensure the validity of subsequent analysis, since assessing the accuracy of recovering the model parameters serves as a prerequisite before evaluating relative model fit.

**Diagnostic Classification Models**

A DCM is a psychometric model which models the probability of answering an item correctly by mastery and non-mastery of a set of attributes (or skills; discrete, ordinal-valued latent variables) that the item measures. Each respondent $e$ has a set of attributes represented as binary indicators of mastery (1) or non-mastery (0), denoted by $\alpha_e = [\alpha_{e1}, \alpha_{e2}, \ldots, \alpha_{eA}]$, where $A$ represents the total number of attributes evaluated in the study. Each item can measure one or more attributes, and the relationship between items $i$ and attributes is represented by a set of Q-matrix entries, denoted as $q_i = [q_{i1}, q_{i2}, \ldots, q_{iA}]$. In the Q-matrix, if an item measures an attribute, the corresponding Q-matrix element takes a value of one, if an item does not measure an attribute, the element takes a value of zero. Each DCM makes different assumptions about relationships between attributes that affect the probability of correctly answering a question. The difference of assumptions is expressed in different equations under the same Q-matrix and respondents' attributes.

A non-compensatory (or conjunctive) assumption considers that all attributes that the item measures should be mastered to answer the item correctly. The deterministic input, noisy-and-gate (DINA) model is a representative model of non-compensatory approach. On the other hand, compensatory assumptions consider that mastered attributes can compensate for non-mastered attributes to increase the probability to answer correctly. The compensatory reparameterized unified model (CRUM) is a compensatory model. When an item is associated with two attributes, the probability of a correct response increases when both required attributes are mastered compared to mastering only one (e.g., Henson et al., 2009). In contract, the log-linear cognitive diagnostic model (LCDM) is a general model that can become either compensatory or non-compensatory depending on the pattern of estimated main effects and interaction item parameters. In other words, the LCDM is flexible enough to show both models based on two different assumptions as Equation 1.

$$P(X_{ie} = 1 \mid \alpha_e) = \frac{\exp(\lambda_{i,0} + \lambda_i^T h(\alpha_e, q_i))}{1 + \exp(\lambda_{i,0} + \lambda_i^T h(\alpha_e, q_i))} \tag{1}$$

Here, $\lambda_{i,0}$ is the intercept which is the logit of a correct response when the examinee has not mastered all attributes examined by item $i$. $\lambda_i$ is the item $i$'s vector of possible main effects of attributes and interaction parameters of the attributes. $h(\alpha_e, q_i)$ is a vector-valued function returning linear combinations of possible attribute profiles when combined with the Q-matrix. For example, when item 1 measures $\alpha_2$ and $\alpha_3$, ($q_1 = [0, 1, 1]$), it can be indicated as Equation 2. $\lambda_{1,1(2)}$ and $\lambda_{1,1(3)}$ are the main effects of $\alpha_{e2}$ and $\alpha_{e3}$ each, and $\lambda_{1,2(2,3)}$ is the interaction effect of both of $\alpha_{e2}$ and $\alpha_{e3}$.

$$P(X_{1e} = 1 \mid \alpha_{e2}, \alpha_{e3}) = \frac{\exp(\lambda_{1,0} + \lambda_{1,1(2)}\alpha_{e2} + \lambda_{1,1(3)}\alpha_{e3} + \lambda_{1,2(2,3)}\alpha_{e2}\alpha_{e3})}{1 + \exp(\lambda_{1,0} + \lambda_{1,1(2)}\alpha_{e2} + \lambda_{1,1(3)}\alpha_{e3} + \lambda_{1,2(2,3)}\alpha_{e2}\alpha_{e3})}. \tag{2}$$

If the examinee's attribute set is $\alpha_e = [1, 1, 0]$, then the probability of answering item 1 correctly is as Equation 3.

$$P(X_{1e} = 1 \mid \alpha_{e2}, \alpha_{e3}) = \frac{\exp(\lambda_{1,0} + \lambda_{1,1(2)})}{1 + \exp(\lambda_{1,0} + \lambda_{1,1(2)})}. \quad (3)$$

Within the context of the LCDM, our research employed two models subsumed by the LCDM, specifically the DINA and CRUM. These models can be expressed by imposing constraints on the parameters of the LCDM. Specifically, in case of the non-compensatory DINA model, Equation 4 shows the probability that examinee $e$ answers item 1 correctly. The main effects of $\alpha_{e2}$ and $\alpha_{e3}$ are omitted, differing from Equation 2 of the LCDM. Only the interaction effect of $\alpha_{e2}$ and $\alpha_{e3}$ remains. This reflects the hypothesis that all attributes associated with the item are required to be mastered by examinees to have a high probability of a correct response to the item.

$$P(X_{1e} = 1 \mid \alpha_{e2}, \alpha_{e3}) = \frac{\exp(\lambda_{1,0} + \lambda_{1,2(2,3)}\alpha_{e2}\alpha_{e3})}{1 + \exp(\lambda_{1,0} + \lambda_{1,2(2,3)}\alpha_{e2}\alpha_{e3})} \quad (4)$$

In contrast, the CRUM retains the two main effects while excluding the interaction effect from Equation 2, as demonstrated by Equation 5. The CRUM is compensatory in that lack of mastery for attribute $\alpha_{e2}$ can be compensated for by mastery of attribute $\alpha_{e3}$.

$$P(X_{1e} = 1 \mid \alpha_{e2}, \alpha_{e3}) = \frac{\exp(\lambda_{1,0} + \lambda_{1,1(2)}\alpha_{e2} + \lambda_{1,1(3)}\alpha_{e3})}{1 + \exp(\lambda_{1,0} + \lambda_{1,1(2)}\alpha_{e2} + \lambda_{1,1(3)}\alpha_{e3})} \quad (5)$$

**Model Fit Indices**

Three relative fit indices DIC, WAIC, and PSIS-LOO were compared in this study. The DIC is a partially Bayesian predictive accuracy measure because it measures the predictive accuracy of the fitted model to data are defined as $-2\log p(y|\hat{\theta}_{Bayes})$ using point estimates ($\hat{\theta}$) instead of posterior distributions $p_{post}(\theta)$ (Gelman et al., 2014).

$$DIC = -2\log p(y|\hat{\theta}_{Bayes}) + 2p_{DIC} \quad (6)$$

The penalty term, denoted as the effective number of parameters ($p_{DIC}$), represents the difference between the mean deviance ($\overline{D(\theta)}$) and the deviance of the means ($D(\bar{\theta})$) in Equation 7, as defined by Spiegelhalter et al. (2002). $\bar{\theta}$ for an attribute is a probability, which is not part of the space of the attribute itself in the DCM context. However, it is worth noting that $p_{DIC}$ may assume a negative value in cases where the posterior mean significantly deviates from the mode.

$$p_{DIC} = 2\left(\log p(y|\hat{\theta}_{Bayes}) - \frac{1}{S}\sum_{s=1}^{S}\log p(y|\theta^s)\right) \tag{7}$$

The WAIC is a fully Bayesian approach to assess the out-of-sample predictive accuracy of a model, similar to DIC. WAIC comprises two components: the predictive accuracy term and the penalty term. The predictive accuracy is determined by the log pointwise predictive density ($lpd$), which is calculated by evaluating the model fit to the data across all individuals using cross-validation. On the other hand, the penalty term ($\hat{p}_{waic}$) corresponds to the effective number of parameters in the model. The $lpd$ considers the entire posterior distribution, allowing for a comprehensive assessment of the model's predictive accuracy. In practice, the computed $lpd$ ($\widehat{lpd}$) is obtained by averaging the draws from the posterior distribution $p_{post}(\theta)$ in $lpd$ in Equation 8, yielding an estimate of the model's predictive accuracy.

$$\widehat{lpd} = \sum_{i=1}^{n}\log\left(\frac{1}{S}\sum_{s=1}^{S}p(y_i|\theta^s)\right) \tag{8}$$

WAIC is an alternative way to estimate the expected log pointwise predictive density and can be defined as

$$\widehat{elpd}_{waic} = \widehat{lpd} - \hat{p}_{WAIC} \tag{9}$$

where $\hat{p}_{waic}$ can be interpreted as a sum over the posterior variances of the log predictive density across individual data points as Equation 10. $\hat{p}_{WAIC}$ is the estimated effective number of parameters as a bias correction.

$$\hat{p}_{WAIC} = \sum_{i=1}^{n} V_{s=1}^{S}\left(\log p(y_i|\theta^s)\right) \tag{10}$$

The expected log pointwise predictive density of LOO estimated with PSIS is

$$\widehat{elpd}_{psis-loo} = \sum_{i=1}^{n} \log \left(\frac{\sum_{s=1}^{S} w_i^s p(y_i|\theta^s)}{\sum_{s=1}^{S} w_i^s}\right) \tag{11}$$

where $w_i^s$ is a vector of weights, s = 1, …, S, for each data point *i*. Vehtari, Gelman, & Gabry (2017) found that while WAIC and PSIS-LOO are asymptotically equivalent, PSIS-LOO exhibits greater stability when weak priors are used in estimation.

**Method**

This study aims to investigate the performance of two Bayesian relative model fit indices, WAIC and PSIS-LOO, in comparison to DIC for the purpose of selecting the appropriate generating models within the family of LCDM. The performance of model fit indices focused on their ability to accurately identify the generating model as observed through the number of selections by the estimation model. Before presenting the accuracy of the model fit, we assessed the estimation accuracy to ensure the validity of the model fit results by examining item recovery, bias, RMSE, and classification accuracy rates, including both full profile and marginal classification rates. The result tables are presented in the Appendix.

*Simulation Study Design*

This research roughly adopted the simulation design of Kunina-Habenicht et al. (2012) and Sen & Bradshaw (2017). We manipulated with five factors: generating model (LCDM, DINA, and CRUM), the number of respondents (500, 1000, and 2000), item quality (medium and high), estimation model (LCDM, DINA, and CRUM), prior information level (uninformative and informative). Replication for each condition was 25 times.

*Constant Factors*

For all conditions, the number of attributes was three, so there were eight possible permutations of attribute statuses. Three attributes were used for this simulation study as they provide enough information to show how these methods generalize to larger numbers of attributes. The test length was fixed at 28 items. All profiles were measured with four items each, except for Profile 1, which did not include any attributes. Thus, each attribute appearing across four profiles, was measured with a total of 16 items. The tetrachoric correlations between each attribute was set to .700. The base rates (i.e., the proportion of mastery for each attribute within the population) for all three attributes were set to .504 to restrict the effect of probability of mastery. The value for the Table 1 is adopted from the Sen & Bradshaw (2017).

**Table 1.** *Profiles and base rate*

| Profile | Attribute 1 | Attribute 2 | Attribute 3 | Proportion |
|---|---|---|---|---|
| 1 | 0 | 0 | 0 | 0.293 |
| 2 | 0 | 0 | 1 | 0.075 |
| 3 | 0 | 1 | 0 | 0.075 |
| 4 | 0 | 1 | 1 | 0.054 |
| 5 | 1 | 0 | 0 | 0.075 |
| 6 | 1 | 0 | 1 | 0.054 |
| 7 | 1 | 1 | 0 | 0.054 |
| 8 | 1 | 1 | 1 | 0.321 |
| Base rate | 0.504 | 0.504 | 0.504 | |

*Manipulated Factors*

**Data generation and estimation model.** Response data were generated based on the LCDM, DINA model, and CRUM. Also, these generated data were estimated by the LCDM, DINA, and CRUM. The LCDM is a more highly parameterized model than the other two variants, DINA model and CRUM. Some model fit indices tend to favor highly parameterized models, so when the generating model is the variant of LCDM, we can verify the performance of indices.

**Number of examinees.** The number of examinees for an item is 500, 1,000, and 2,000. To investigate whether Bayesian estimation performs properly at small level of sample sizes, 500 respondents were included in the design.

**Item quality.** The quality of items was set to two levels, medium and high. When the quality of item is high, items can highly discriminate between respondents classified as being masters as compared to those classified as being non-masters. High quality items have a bigger difference in the probability of a correct response conditional on a respondent's mastery status than do less discriminating, lower quality items. The item parameters were set based on the gap between masters and non-masters.

**Table 2**. *Item quality based on the probability of correct response for an item*

| Number of mastered attributes | Correct Response Probability | | Item parameter | Value | |
|---|---|---|---|---|---|
| | Medium | High | | Medium | High |
| 0 | 0.25 | 0.12 | Intercept | -1.10 | -2.0 |
| 1 | 0.55 | 0.50 | Main | 1.30 | 2.0 |
| 2 | 0.85 | 0.95 | Two-way | 0.23 | 1.0 |
| 3 | 0.99 | 0.99 | Three-way | 3.40 | 0.1 |

*Note*. Two-way = Two-way interaction. Three-way = Three-way interaction, Medium = Medium quality, High = High quality

As shown in Table 2, the probability of a correct response between respondents who have mastered no attributes and those who have mastered all attributes is .74 for a medium-quality item and .87 for a high-quality item. When items measure a single attribute, the LCDM, DINA model, and CRUM use the value of one main effect term from Table 2. However, when items measure two or more attributes, the LCDM incorporates the values of the intercept, main effects, and interaction terms. The DINA model uses only the intercept and interaction terms, while CRUM relies on the intercept and main effect terms.

**Prior information level.** The prior information levels for the item parameters were set as either informative or uninformative, with the prior distributions following a multivariate normal distribution. The hyperparameters for the informative prior were set with a mean of 0, variance of 5. This selection was based on previous research, particularly the tutorials by Jiang and Carter (2019) and Zhan et al. (2019), which focused on the analyzing DCMs using Stan and JAGS. Both studies set the normal distribution with a mean of 0. Jiang and Carter (2019) used a variance of 25, while Zhan et al. (2019) set a variance of 0.25. To balance between these two approaches, we selected a variance of 5. The uninformative prior distribution's hyperparameters were mean of 0, variance of 1000. The covariance of the distribution for both the informative and uninformative priors were set to 0.

*Estimation Methods*

Bayesian estimation was used for estimating the parameters of LCDM, DINA model, and CRUM. Bayesian posterior parameter estimates were obtained using Markov Chain Monte Carlo (MCMC) with Gibbs sampling algorithms implemented in the **blatent** (Templin, 2020) R package (R Core Team, 2022). The four Markov chains were used, with each chain running between 2,000 and 12,000 iterations. While the number of iterations per chain was consistent within each replication, it varies across different replications. The first 1,000 to 4,000 iterations were discarded as burn-in, while the remaining 1,000 to 8,000 iterations were used for sampling the posterior distribution. When the model parameters failed to converge, both the burn-in and sampling iterations were increased simultaneously to ensure stable estimation, and re-estimated. For instance, to obtain 8,000 sampling iterations, a burn-in period of 4,000 iterations was required. The latent class membership probabilities follow Dirichlet (1) distribution as prior distribution.

*Evaluation Criteria*

A diagnosis of convergence of MCMC estimation was conducted by Gelman-Rubin Potential Scale Reduction Factor (PSRF denoted with $\hat{R}$). As the maximum of $\hat{R}$ is recommended to be under 1.1 (Gelman et al., 2014), the number of burn-in and sampling iteration was increased when the $\hat{R}$ was higher than 1.1. To check the relative model fit, DIC and WAIC were computed in the **blatent** package and PSIS-LOO was computed with the **loo** (Vehtari et al., 2022) package using the marginal log likelihood generated in the **blatent** package.

To validate the performance of model fit indices, the parameter recovery was examined first. Item parameter recovery was evaluated using both bias and the root mean square error (RMSE). There are two types of classification accuracy used to assess the recovery of an examinee's ability: attribute profile classification rate and marginal attribute classification rate. The attribute profile classification rate shows the ratio of all three attributes recovered, whereas the marginal attribute classification rate represents the ratio of a recovered single attribute.

**Results**

*Item Parameter Recovery*

Figure A1 and Figure A2 in Appendix shows bias for each item parameter by condition in a box plot in the order of intercept, main effect, two-way interaction, and three-way interaction, with Figure A1 representing results under the informative prior condition and Figure A2 showing results under the uninformative prior condition. These results are comparable with previous studies showing decreasing accuracy of complex effects (Kunina-Habenicht et al., 2012). Based on the two figures, bias values for the intercept parameter ranged from -0.2 to 0.25 showing that parameter estimates for the intercept were recovered well for most of the models. Bias values for the main effects are mostly within -0.25 to 0.25, and those of interaction effects were mostly ranged from -3.5 to 1.0. Comparing each parameter between models, the LCDM was the best model to recover the intercept and main effect parameters with low quality items. DINA was the best model to recover two- and three-way interactions. The RMSE for item parameter estimates for each condition are represented in Figure A3 and Figure A4. The patterns of RMSE of intercept (0.05–0.35), main effect (0.05–0.6), and two-way interaction (0.0–1.5) were similar with previous studies (Sen & Bradshaw, 2017). The RMSE of three-way interaction indicated large RMSE ranged from 0.00 to 3.25 as expected. LCDM was the most well-recovered model for the intercept and main effects, and DINA was the most well-recovered model for interaction as predicted due to the complexity of model, which was consistent with the results of the bias (Kunina-Habenicht et al., 2012).

*Classification Accuracy*

Table A1 and Table A2 shows the profile classification rate (classification into one of the eight attribute profiles, jointly) and marginal attribute classification rate (classification for each attribute, marginally). The classification rate for each generated model was unlikely to be similar across the six conditions. Based on the two figures, the profile classification rate ranged from 70.11% to 91.39% for the LCDM, 56.42% to 60.20% for the DINA, and 66.94% to 86.74% for the CRUM as expected (Galeshi, 2012; Huang & Wang, 2014; Sen & Bradshaw, 2017). The LCDM showed the highest profile classification accuracy under all conditions and the CRUM was the next highest. Additionally, the profile classification rates did not increase with sample size, so the sample size did not significantly impact the profile accuracy. Marginal classification rates for all conditions exhibited over 80% showing a similar pattern with profile classification accuracy rate.

*Model Selection Indices*

Table 3 presents the performance of three model fit indices—DIC, WAIC and PSIS-LOO—under the informative prior condition. The values in Table 3 indicate the number of each model selected as the best model among 25 iterations, considering different conditions such as the generating model, sample size, and item quality. To compare the performance of three different indices based on the characteristics of the generating model, the table is divided into three columns corresponding to each generating model. For instance, the top left corner of the table corresponded to the condition with a sample size of 500, medium item quality, and the LCDM as the generating model. the DIC selected the LCDM as the best fit 21 times, while the DINA model was not selected at all, and the CRUM was chosen as the best fit 4 times.

In all conditions, DIC, WAIC, and PSIS-LOO consistently selected the correct model with the highest frequency, and there were occasional inaccuracies where the wrong model was chosen. When the generating model was LCDM, all indices selected the LCDM 25 times, except for medium-quality conditions with sample sizes of 500 and 1,000, where the CRUM was incorrectly selected. Notably, the accuracy rate was higher for the 1,000-sample size compared to the 500-sample size under medium-quality conditions, as expected due to the larger sample size resulting in more accurate estimates. Moreover, DIC selected the correct model more frequently than WAIC and PSIS-LOO.

When the DINA model was the generating model, all indices selected the best estimation model the same number of times, regardless of whether the estimation model was correct or incorrect. Unlike when the generating model was LCDM, when all indices selected the wrong estimation model, it was under the condition of high-quality conditions. Under high-quality conditions, CRUM was selected 6 times with a sample size of 500 and 4 times with a sample size of 1,000. Lower accuracy at higher quality item is the expected result in DINA model (Huang & Wang, 2014).

**Table 3.** *Model selection based on three model fit indices under informative prior*

| Sample size | Item quality | Est. model | Generating model | | | | | | | | |
|---|---|---|---|---|---|---|---|---|---|---|---|
| | | | LCDM | | | DINA | | | CRUM | | |
| | | | DIC | WAIC | PSIS LOO | DIC | WAIC | PSIS LOO | DIC | WAIC | PSIS LOO |
| 500 | Medium | LCDM | 21 | 19 | 19 | 0 | 0 | 0 | 0 | 0 | 0 |
| | | DINA | 0 | 0 | 0 | 25 | 25 | 25 | 0 | 0 | 0 |
| | | CRUM | 4 | 6 | 6 | 0 | 0 | 0 | 25 | 25 | 25 |
| | High | LCDM | 25 | 25 | 25 | 0 | 0 | 0 | 0 | 0 | 0 |
| | | DINA | 0 | 0 | 0 | 19 | 19 | 19 | 0 | 0 | 0 |
| | | CRUM | 0 | 0 | 0 | 6 | 6 | 6 | 25 | 25 | 25 |
| 1000 | Medium | LCDM | 23 | 22 | 22 | 0 | 0 | 0 | 0 | 0 | 0 |
| | | DINA | 0 | 0 | 0 | 25 | 25 | 25 | 0 | 0 | 0 |
| | | CRUM | 2 | 3 | 3 | 0 | 0 | 0 | 25 | 25 | 25 |
| | High | LCDM | 25 | 25 | 25 | 0 | 0 | 0 | 0 | 0 | 0 |
| | | DINA | 0 | 0 | 0 | 21 | 21 | 21 | 0 | 0 | 0 |
| | | CRUM | 0 | 0 | 0 | 4 | 4 | 4 | 25 | 25 | 25 |
| 2000 | Medium | LCDM | 25 | 25 | 25 | 0 | 0 | 0 | 0 | 0 | 0 |
| | | DINA | 0 | 0 | 0 | 25 | 25 | 25 | 0 | 0 | 0 |
| | | CRUM | 0 | 0 | 0 | 0 | 0 | 0 | 25 | 25 | 25 |
| | High | LCDM | 25 | 25 | 25 | 0 | 0 | 0 | 0 | 0 | 0 |
| | | DINA | 0 | 0 | 0 | 25 | 25 | 25 | 0 | 0 | 0 |
| | | CRUM | 0 | 0 | 0 | 0 | 0 | 0 | 25 | 25 | 25 |

*Note.* Est. model = estimation model; LCDM = log-linear cognitive diagnostic model; DINA = deterministic input, noisy "and" gate; CRUM = compensatory reparameterized unified model, DIC = deviance information criterion, WAIC = widely applicable information criterion, PSIS-LOO = pareto-smoothed importance sampling

All indices when the generating model was the CRUM exhibited the better performance compared to when the generating models were the LCDM and DINA models, particularly at a sample size of 500. All indices perfectly selected the CRUM as the best-fitted model.

Table 4 presents the performance of the three model fit indices — DIC, WAIC, and PSIS-LOO — under uninformative prior conditions. The uninformative prior distribution results followed a similar pattern to those observed under the informative prior condition. In all conditions, DIC, WAIC, and PSIS-LOO consistently selected the correct model with the highest frequency, though inaccuracies occasionally occurred, with incorrect models being chosen.

**Table 4.** *Model selection based on three model fit indices under uninformative prior*

| Sample size | Item quality | Est. model | Generating model ||||||||| 
| | | | LCDM ||| DINA ||| CRUM |||
| | | | DIC | WAIC | PSIS LOO | DIC | WAIC | PSIS LOO | DIC | WAIC | PSIS LOO |
|---|---|---|---|---|---|---|---|---|---|---|---|
| 500 | Medium | LCDM | 18 | 18 | 18 | 0 | 0 | 0 | 0 | 0 | 0 |
| | | DINA | 0 | 0 | 0 | 25 | 25 | 25 | 0 | 0 | 0 |
| | | CRUM | 7 | 7 | 7 | 0 | 0 | 0 | 25 | 25 | 25 |
| | High | LCDM | 25 | 25 | 25 | 0 | 0 | 0 | 0 | 0 | 0 |
| | | DINA | 0 | 0 | 0 | 17 | 17 | 17 | 0 | 0 | 0 |
| | | CRUM | 0 | 0 | 0 | 8 | 8 | 8 | 25 | 25 | 25 |
| 1000 | Medium | LCDM | 24 | 22 | 22 | 0 | 0 | 0 | 0 | 0 | 0 |
| | | DINA | 0 | 0 | 0 | 25 | 25 | 25 | 0 | 0 | 0 |
| | | CRUM | 1 | 3 | 3 | 0 | 0 | 0 | 25 | 25 | 25 |
| | High | LCDM | 25 | 25 | 25 | 0 | 0 | 0 | 1 | 1 | 1 |
| | | DINA | 0 | 0 | 0 | 23 | 23 | 23 | 0 | 0 | 0 |
| | | CRUM | 0 | 0 | 0 | 2 | 2 | 2 | 24 | 24 | 24 |
| 2000 | Medium | LCDM | 25 | 25 | 25 | 0 | 0 | 0 | 1 | 0 | 0 |
| | | DINA | 0 | 0 | 0 | 25 | 25 | 25 | 0 | 0 | 0 |
| | | CRUM | 0 | 0 | 0 | 0 | 0 | 0 | 24 | 25 | 25 |
| | High | LCDM | 25 | 25 | 25 | 0 | 0 | 0 | 0 | 0 | 0 |
| | | DINA | 0 | 0 | 0 | 25 | 25 | 25 | 0 | 0 | 0 |
| | | CRUM | 0 | 0 | 0 | 0 | 0 | 0 | 25 | 25 | 25 |

*Note.* Est. model = estimation model; LCDM = log-linear cognitive diagnostic model; DINA = deterministic input, noisy "and" gate; CRUM = compensatory reparameterized unified model, DIC = deviance information criterion, WAIC = widely applicable information criterion, PSIS-LOO = pareto-smoothed importance sampling

When the generating model was LCDM, DIC, WAIC, and PSIS-LOO demonstrated similar performance, with the correct model being selected across most conditions. For instance, under medium-quality conditions with a sample size of 500, the LCDM was selected 18 times, while CRUM was selected 7 times, consistent across all indices. At a sample size of 1,000 under medium-quality conditions, the LCDM was selected 24 times by DIC, while WAIC and PSIS-LOO selected the LCDM 22 times. The CRUM was selected 3 times by WAIC and PSIS-LOO under this condition. Under higher sample sizes of 1,000 and 2,000 and high-quality item condition, the correct model selection frequency increased, with the LCDM being selected perfectly across all indices.

When DINA was the generating model, all three indices consistently identified the DINA model as the best fit under medium-quality conditions, regardless of sample size. For example, at a sample size of 500 under medium-quality conditions, the DINA model was selected 25 times, showing high accuracy across all indices. However, when item quality improved, such as in high-quality conditions with a sample size of 500 and 1,000, the CRUM was occasionally selected, consistent with findings under informative priors.

Lastly, when the CRUM was the generating model, all indices generally performed well. However, a notable difference from the informative prior results was observed under the 1,000 sample size and high-quality conditions, where all three indices selected the LCDM once. Additionally, under medium-quality conditions with a sample size of 2,000, the DIC selected the LCDM once, whereas the WAIC and PSIS-LOO consistently identified the CRUM as the best-fitting model. In all but two conditions, the CRUM was consistently identified as the best-fitting model, maintaining its overall performance across conditions despite these anomalies.

**Discussion**

This paper introduces the performance of WAIC and PSIS-LOO in comparison to DIC, which is widely used within the framework of DCMs. The three model fit indices generally favored the generating model, but there were occasional inconsistencies with WAIC and PSIS-LOO when the LCDM or DINA model was the generating model, and with DIC when the CRUM was the generating model. The study suggests considering multiple fit indices when dealing with complex models, diverse item qualities and limited sample sizes.

The simulation results indicated that the item parameters were recovered as expected with the accuracy varying depending on the model. In terms of the LCDM, the intercept and main effect of the LCDM were well recovered, while the interaction term was not recovered as accurately as the intercept and main effects. This is likely due to the interactions being more challenging to estimate due to their complexity in the LCDM (Sen & Bradshaw, 2017). On the other hand, the DINA model recovered the two-way interaction and three-way interaction better than the intercept and main effect.

All model fit indices consistently selected the generating model as the best model across all conditions with the highest frequency, suggesting that the three indices generally performed well for DCMs. A few instances of incorrect models were selected under specific conditions. When the generating model was the LCDM under the condition of medium item quality and sample sizes of 500 and 1,000, three indices selected the CRUM as the best model for both prior information levels. We attribute this inaccuracy to the difficulty in recovering the interaction effects (Sen & Bradshaw, 2017). The bias analysis in Figure A1 showed that the three-way interaction was estimated as nearly zero under the LCDM model because the true value of the interaction term was 3.4, resulting in a bias value around -3. This suggests that the CRUM which excludes the interaction term present in the LCDM might have been selected as the best model due to the interaction term's estimated value being zero.

While all three indices demonstrated favorable performance, both WAIC and PSIS-LOO selected the best model at the equivalent number of times across all conditions (Luo & Al-Harbi, 2017). This was expected results because WAIC asymptotically approximates LOO (Vehtari et al., 2017). Also, it is noteworthy that when the generating model was the LCDM, WAIC and PSIS-LOO showed a higher proportion of selecting CRUM compared to DIC under the condition of medium item quality and a sample size of 500 and 1,000. This finding diverges from previous studies conducted with unidimensional IRT models. The earlier study demonstrated a high performance of WAIC and LOO when the generating model was the 3PL IRT model which had more parameters compared to the 2PL and 1PL IRT models (Luo & Al-Harbi, 2017). This discrepancy may stem from WAIC's reliance on data partitioning, which could pose challenges with structured models (Gelman et al., 2014). Therefore, the assumption of dimensionality might influence the performance of WAIC. Based on these results, it is recommended to assess the performance of WAIC and PSIS-LOO on higher-order DCMs in the future research.

On the other hand, the DIC had a higher rate to select the correct model compared to the WAIC and PSIS-LOO when the generating model was the LCDM under the condition of medium item quality and 500 and 1,000 sample size. Although the DIC selected the correct model one to two times more out of 25 repetitions than the WAIC and PSIS-LOO, it consistently made more accurate model selections. Therefore, it is recommended to use DIC instead of WAIC and PSIS-LOO when estimating with models that have more parameters, limited sample size, and low item quality regardless of prior information level, all of which make the estimation challenging. Alternatively, the use of more than one index may be needed to ensure more accurate model comparisons.

      Contrastingly, when the generating model was CRUM, the three indices consistently and accurately selected the correct model under the informative prior condition. However, under the uninformative prior with a sample size of 2000 and medium-quality item conditions, DIC selected LCDM as the best-fitting model. This finding is aligned with the previous research that DIC tends to favor more complex models (Fujimoto & Falk, 2022). Thus, the two indices, WAIC and PSIS-LOO, emerge as reliable methods for selecting the parsimonious model within the DCM context.

      Consequently, under all conditions, WAIC and PSIS-LOO performed similarly to DIC. While the recovery of parameters was as expected, achieving more accurate parameter recovery would further enhance the validity of determining the performance of the three relative fit indices. Furthermore, since the three indices exhibited varying rates of incorrect model selection depending on the conditions, future research should consider adding different manipulated factors (e.g., Q-matrix misspecification, base rate) to determine the performance of all indices.

# Appendix

## Appendix A. Item parameter recovery and classification accuracy

**Figure A1.** *Bias for item parameter estimates for each condition under informative prior*

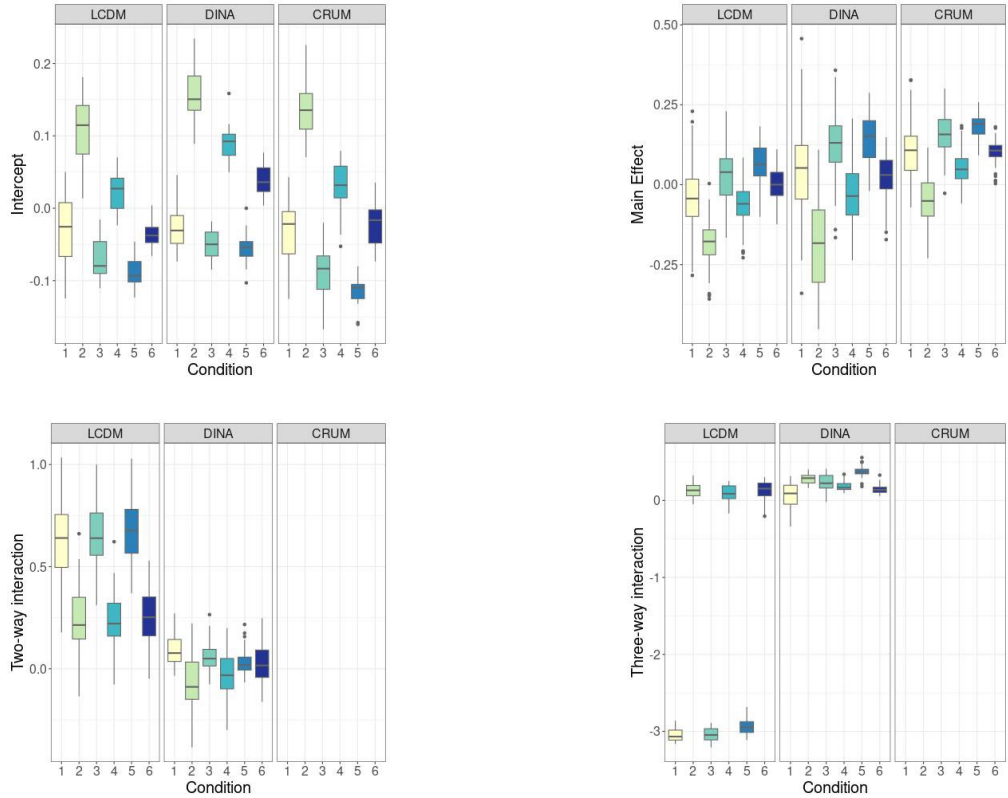

*Note*. Condition 1: Sample size = 500, Item quality = Medium
Condition 2: Sample size = 500, Item quality = High
Condition 3: Sample size = 1000, Item quality = Medium
Condition 4: Sample size = 1000, Item quality = High
Condition 5: Sample size = 2000, Item quality = Medium
Condition 6: Sample size = 2000, Item quality = High

**Figure A2.** *Bias for item parameter estimates for each condition under uninformative prior*

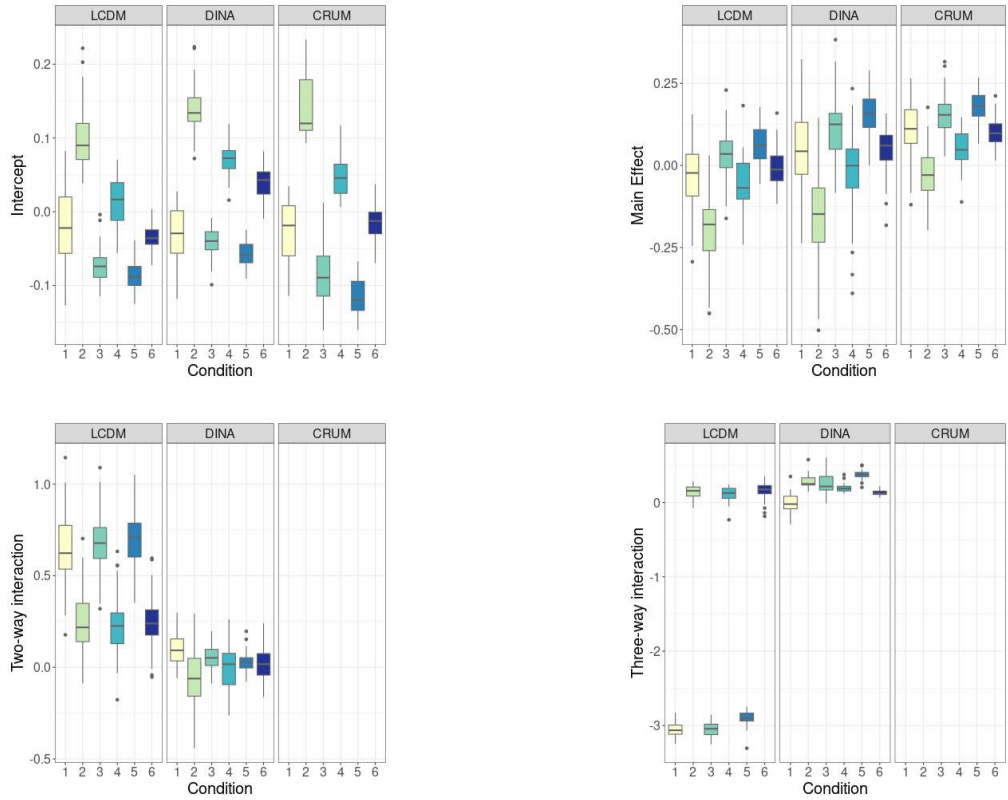

*Note*. Condition 1: Sample size = 500, Item quality = Medium
Condition 2: Sample size = 500, Item quality = High
Condition 3: Sample size = 1000, Item quality = Medium
Condition 4: Sample size = 1000, Item quality = High
Condition 5: Sample size = 2000, Item quality = Medium
Condition 6: Sample size = 2000, Item quality = High

**Figure A3.** *RMSE for item parameter estimates for each condition under informative prior*

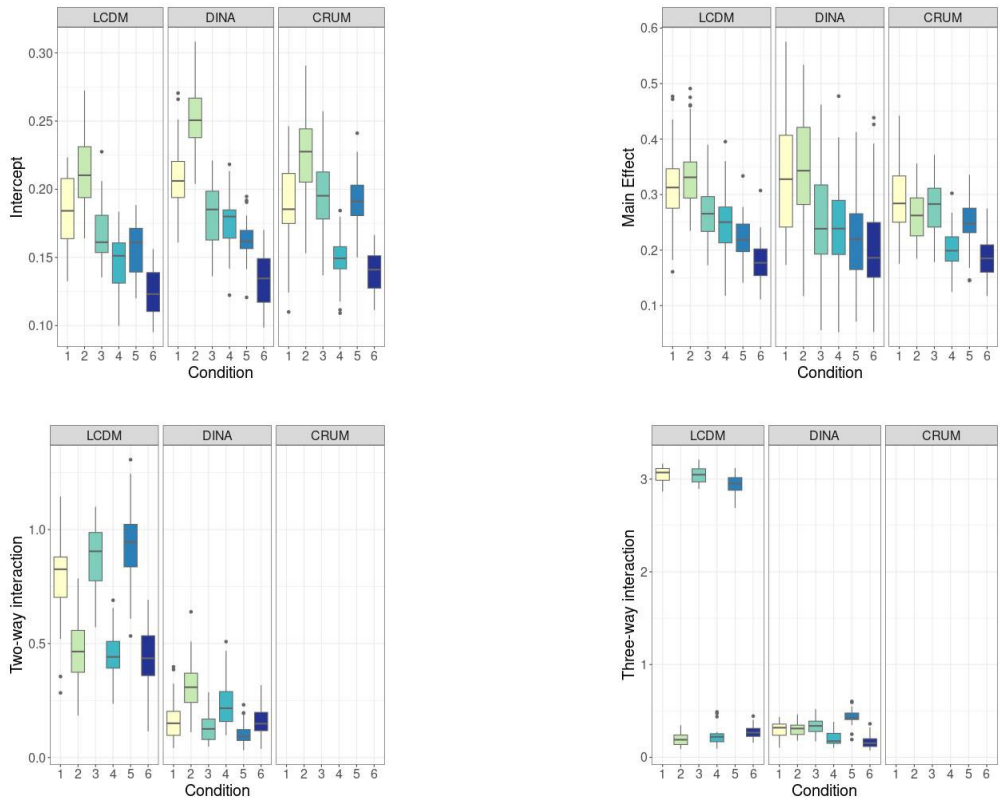

*Note.* Condition 1: Sample size = 500, Item quality = Medium
Condition 2: Sample size = 500, Item quality = High
Condition 3: Sample size = 1000, Item quality = Medium
Condition 4: Sample size = 1000, Item quality = High
Condition 5: Sample size = 2000, Item quality = Medium
Condition 6: Sample size = 2000, Item quality = High

**Figure A4.** *RMSE for item parameter estimates for each condition under uninformative prior*

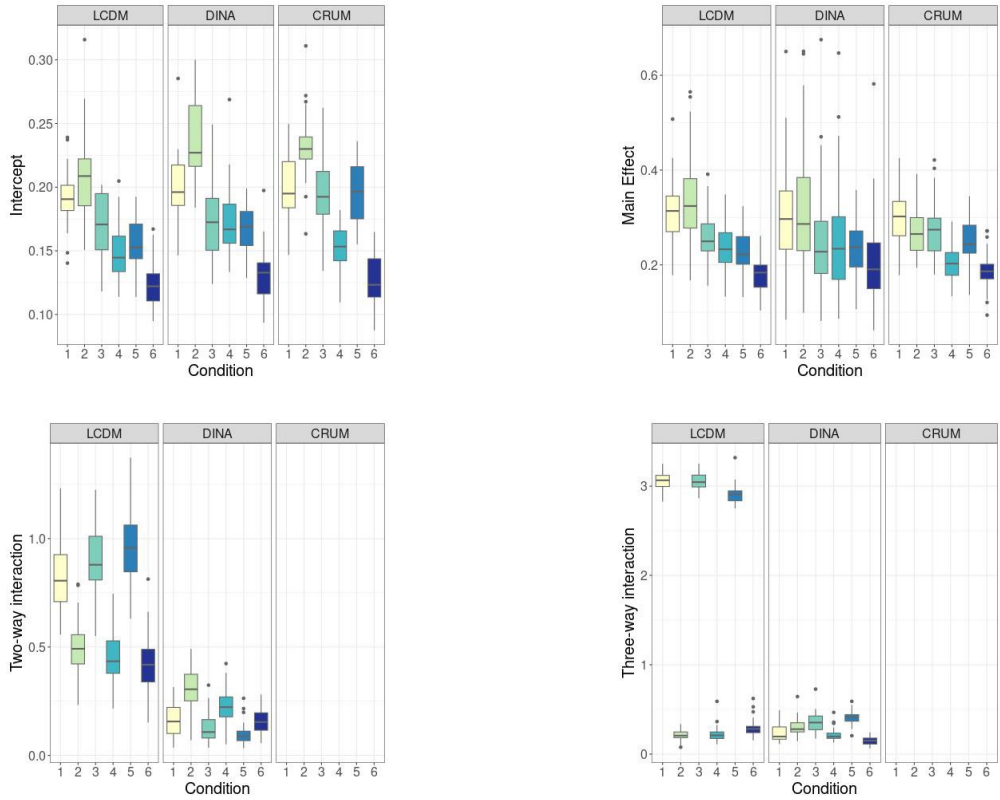

*Note.* Condition 1: Sample size = 500, Item quality = Medium
Condition 2: Sample size = 500, Item quality = High
Condition 3: Sample size = 1000, Item quality = Medium
Condition 4: Sample size = 1000, Item quality = High
Condition 5: Sample size = 2000, Item quality = Medium
Condition 6: Sample size = 2000, Item quality = High

**Table A3.** *Mean of classification accuracy rates for each condition under informative prior*

| N | Quality | Profile | | | Marginal | | |
|---|---------|---------|------|------|----------|------|------|
|   |         | LCDM | DINA | CRUM | LCDM | DINA | CRUM |
| 500 | Medium | 0.7029 | 0.5642 | 0.6790 | 0.8927 | 0.8071 | 0.8869 |
|     | High   | 0.9139 | 0.5811 | 0.8622 | 0.9667 | 0.8330 | 0.9508 |
| 1000 | Medium | 0.7074 | 0.5720 | 0.6744 | 0.8972 | 0.8107 | 0.8874 |
|      | High   | 0.9079 | 0.5966 | 0.8674 | 0.9643 | 0.8374 | 0.9521 |
| 2000 | Medium | 0.7011 | 0.5672 | 0.6695 | 0.8959 | 0.8084 | 0.8891 |
|      | High   | 0.9088 | 0.5964 | 0.8660 | 0.9650 | 0.8375 | 0.9526 |

Note. Profile = Attribute Profile Classification, Marginal = Marginal Attribute Classification

**Table A2.** *Mean of classification accuracy rates for each condition under uninformative prior*

| N | Quality | Profile | | | Marginal | | |
|---|---------|---------|------|------|----------|------|------|
|   |         | LCDM | DINA | CRUM | LCDM | DINA | CRUM |
| 500 | Medium | 0.7138 | 0.5693 | 0.6812 | 0.8992 | 0.8115 | 0.8879 |
|     | High   | 0.9135 | 0.5896 | 0.8589 | 0.9669 | 0.8371 | 0.9496 |
| 1000 | Medium | 0.7107 | 0.5688 | 0.6694 | 0.8981 | 0.8092 | 0.8860 |
|      | High   | 0.9100 | 0.5903 | 0.8616 | 0.9660 | 0.8369 | 0.9514 |
| 2000 | Medium | 0.7015 | 0.5755 | 0.6710 | 0.8956 | 0.8109 | 0.8887 |
|      | High   | 0.9087 | 0.6020 | 0.8634 | 0.9651 | 0.8409 | 0.9520 |

Note. Profile = Attribute Profile Classification, Marginal = Marginal Attribute Classification